\documentclass[sigconf]{acmart}

\usepackage{multirow}
\usepackage{colortbl}
\usepackage{bm}
\usepackage{enumitem}
\usepackage{mathtools}

\usepackage{float}
\usepackage{listings}

\definecolor{codegreen}{RGB}{0,128,0}
\definecolor{codeblue}{RGB}{0,0,255}
\definecolor{backgray}{RGB}{248,248,248}
\definecolor{numgray}{gray}{0.4}

\renewcommand\footnotetextcopyrightpermission[1]{}

\begin{document}

\title{SOARS: System One Assisted Reranking for Agentic Search}

\author{Amin Bigdeli}
\orcid{0009-0003-8977-9312}
\affiliation{%
  \institution{University of Waterloo}
  \country{}
}
\email{abigdeli@uwaterloo.ca}

\author{Liye Fu}
\orcid{0000-0001-7989-6839}
\affiliation{%
     \institution{Thomson Reuters Labs}
     \country{}
}
\email{liye.fu@thomsonreuters.com}

\author{Fattane Zarrinkalam}
\orcid{0000-0002-2102-9190}
\affiliation{%
  \institution{University of Guelph}
  \country{}
}
\email{fzarrink@uoguelph.ca}

\author{Joel Stremmel}
\orcid{0009-0002-4540-7982}
\affiliation{%
  \institution{Thomson Reuters Labs}
  \country{}
}
\email{joel.stremmel@thomsonreuters.com}

\author{Elizabeth Lingg}
\orcid{0009-0009-8597-7665}
\affiliation{%
  \institution{Thomson Reuters Labs}
  \country{}
}
\email{elizabeth.lingg@thomsonreuters.com}

\renewcommand{\shortauthors}{Amin Bigdeli et al.}

\begin{abstract}
Deep search agents address complex information needs through iterative retrieval and reasoning, where the relevance of retrieved evidence influences downstream performance. Existing rerankers improve retrieval effectiveness but introduce substantial latency and cost when invoked throughout the search trajectory. We introduce SOARS, a training-free framework leveraging a System One model for fast, calibrated relevance estimation. At each search step, SOARS selects query-relevant passages from long documents and reranks candidates in a single call, jointly considering the original question and agent-generated sub-query. Across six LLM backbones on BrowseComp-Plus and WebShaper, SOARS consistently improves answer accuracy and evidence recall while reducing search calls, end-to-end latency, and inference cost. On BrowseComp-Plus, it improves GPT-5.4 Mini's accuracy from 73.9\% to 84.7\% while reducing search calls from 23 to 14, demonstrating the potential of System One reranking for more effective and efficient deep search. 

\end{abstract}

\maketitle

\section{Introduction}
\label{sec:intro}

\definecolor{citeblue}{HTML}{1A4FB0}

\begin{figure}[t]
  \centering

    \includegraphics[width=0.475\textwidth,
            trim=0.4cm 0cm 0cm 0cm,
        clip
        ]{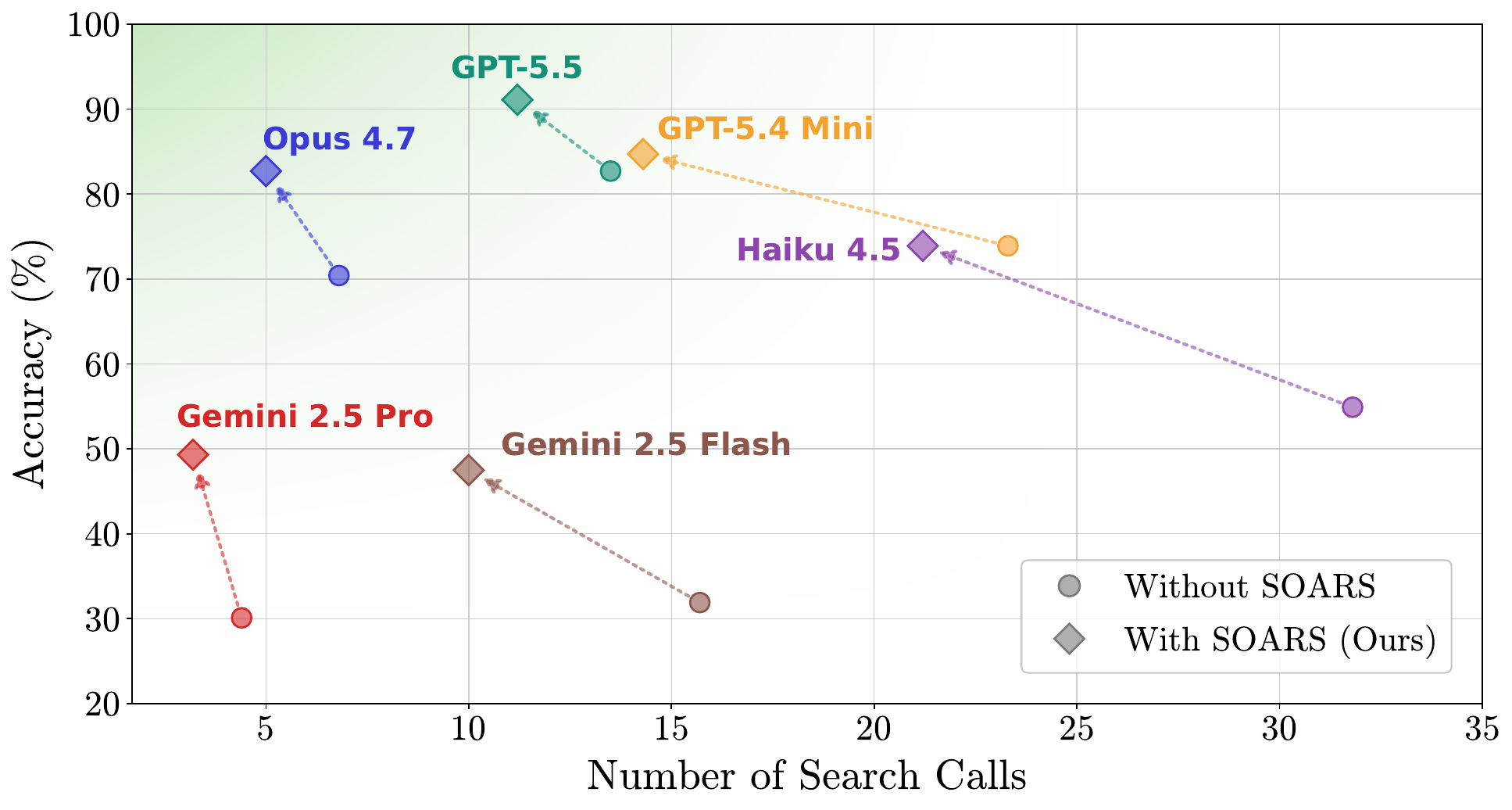}

\vspace{-1em}
\caption{Effect of SOARS on accuracy and search calls of deep search agents across various LLMs on BrowseComp-Plus~\cite{browscompplus}.}

\vspace{-1em}
\label{fig:motivation_figure}
\end{figure}

Deep search agents leverage Large Language Models (LLMs) to address complex information needs through iterative rounds of searching, reading, and reasoning, deciding what to retrieve and read next based on the evidence collected throughout the search process~\cite{react,piserini,huang2025deep,agentic_IR}. This paradigm supports reasoning-intensive tasks whose answers are not contained in any single document but must be assembled from evidence distributed across multiple documents~\cite{browscompplus}. Given the limited interaction budgets of deep search agents, a central challenge in Web search is enabling these agents to surface the most relevant evidence effectively and efficiently.

To provide deep search agents with relevant corpus evidence, prior work has explored several directions. Some approaches improve first-stage retrieval by training dense retrievers for agent-generated sub-queries~\cite{agentir} or adapting lexical retrieval to long documents~\cite{piserini}. Another approach warm-starts the agent by decomposing the question and reranking results from complementary searches to construct an initial evidence context to guide iterative search and reasoning~\cite{question_gambit}. Retrieval-aware agent control has also been explored to mitigate reasoning stagnation and reduce redundant searches by monitoring information coverage and search novelty~\cite{deep_search_agents_stagnate}. However, these approaches focus on retrieval, initialization, or trajectory control rather than reranking evidence throughout the search process.

Reranking first-stage retrieval candidates is a common approach to improving retrieval effectiveness, typically using cross-encoders~\cite{nogueira2019passage,monobert} or LLM-based rerankers that assess relevance through prompting or reasoning~\cite{rankgpt,rankzephyr,rank-r1,rankvicuna}. Recent studies have extended reranking to deep search agents~\cite{meng2026ranking,rerankbeforeyoureason}, demonstrating improvements in retrieval effectiveness and answer accuracy. Within a deep search agent, however, reranking is not a one-time operation but is often invoked at every search call along the trajectory, adding latency without substantially reducing the number of searches. Moreover, existing approaches typically rank candidates solely based on agent-generated sub-queries, overlooking the original question's global information need, and rely on initial document tokens that may omit relevant evidence. Consequently, they improve retrieval effectiveness at the expense of efficiency, without substantially reducing the agent's search effort.

Emerging System One models offer a promising direction for jointly improving retrieval effectiveness and efficiency. Unlike generative LLMs, these models produce structured, calibrated decisions in a single forward pass, enabling fast and cost-efficient inference~\cite{jev_laya,system-one}. Recent work by \citet{jev_as_a_judge} demonstrates that Jev, a System One model, achieves relevance judgments comparable to frontier models at substantially lower latency and cost. These capabilities are particularly valuable for deep search agents, where retrieval and reranking occur repeatedly. By prioritizing relevant evidence at each search step, System One reranking can expose agents to more useful information, potentially reducing search calls, agent reasoning overhead, and cost. However, adapting these models to deep search requires addressing long documents and the interplay between the global information need of the original question and the local information need of agent-generated sub-queries.

To this end, the \textbf{main objective of this paper} is to investigate how lightweight, repeated System One reranking improves the effectiveness and efficiency of deep search agents. Specifically, we address three research questions (RQs). \textbf{(RQ1)} How can System One models be adapted as in-loop rerankers in the context of deep search agents to improve retrieval effectiveness? \textbf{(RQ2)} How does System One reranking affect the agent's downstream answer accuracy and evidence recall? \textbf{(RQ3)} What is its impact on overall agent efficiency in terms of latency and cost?

To investigate these questions, we introduce SOARS (\textbf{S}ystem \textbf{O}ne \textbf{A}ssisted \textbf{R}eranking for Agentic \textbf{S}earch), a training-free framework that integrates System One reranking into deep search agents. SOARS represents long documents using query-relevant passages and reranks candidates in a single call, considering both the original question and the agent-generated sub-query. We evaluate SOARS on two deep search benchmarks, BrowseComp-Plus~\cite{browscompplus} and WebShaper~\cite{webshaper}, across six LLMs of varying families and sizes. Our results show that SOARS achieves retrieval effectiveness comparable to strong LLM-based rerankers while operating one to two orders of magnitude faster. At the agent level, SOARS prioritizes relevant evidence for subsequent reasoning, improving answer accuracy and reducing search calls across all backbones (Figure~\ref{fig:motivation_figure}). Notably, GPT-5.4 Mini's accuracy rises from 73.9\% to 84.7\% while search calls drop from 23 to 14. These improvements also translate into lower latency and cost, demonstrating how efficient, repeated reranking can improve the overall search trajectory.

\section{Proposed Approach}
\label{sec:approach}

\textbf{Problem Formulation.}
\label{sec:formulation}
Given a complex question $Q$ with supporting evidence distributed across corpus $\mathcal{D}$, a deep search agent produces an answer by iteratively searching, reading, and reasoning over retrieved evidence. The agent is driven by a large language model $\mathcal{S}_{\theta}$ acting through a policy $\pi_{\theta}$ in a \textsc{ReAct}-style loop~\cite{react}. At each turn $t$, it emits a reasoning trace $\tau_t$, selects an action $a_t$, and receives an observation $o_t$, forming the interaction trajectory as shown below:

\begin{equation}
    H_T = (\tau_1, a_1, o_1, \dots, \tau_T, a_T, o_T), \qquad (\tau_t, a_t) \sim \pi_{\theta}(\cdot \mid H_{t-1}).
\end{equation}
A search action issues a sub-query $q_t$ to a retriever, which returns the list $L_{q_t}$ of documents retrieved for $q_t$. The agent operates under a limited interaction budget, so the order in which retrieved evidence is presented, and the form in which each document is exposed, determine what the agent ever reads and reasons over. Reranking $L_{q_t}$ can reshape this evidence, but since it is repeated at every search call along the trajectory, it must be not only effective but also efficient to be practical. This work introduces SOARS, a training-free reranker that reranks the candidates returned by every in-loop search before they reach the agent.

\textbf{Deep Search Agent.}
\label{sec:agent}
The agent interacts with the corpus through issuing searches, previewing the returned results, and reading documents in full~\cite{piserini}. At each turn $t$, when the agent selects a search action with sub-query $q_t$, a first-stage retriever $\mathcal{R}$ surfaces documents from $\mathcal{D}$ and returns a ranked list as below:
\begin{equation}
    L_{q_t} = \mathcal{R}(q_t, \mathcal{D}).
\end{equation}
The agent then previews and reads from $L_{q_t}$ to gather evidence for $Q$. Because the agent inspects only the top of this list, the documents placed near its top determine what evidence enters the trajectory. Our objective is to rerank $L_{q_t}$ at each search efficiently in a way that the agent is exposed to more relevant documents.

\textbf{System One Reranking.}
\label{sec:reranking}
SOARS reranks the top $k$ documents of $L_{q_t}$ by preparing a query and a compact representation for each document and scoring them with a System One model. This would let more candidates fit within the model's context, so a large pool is reranked in a single call.

Since an agent-generated sub-query $q_t$ captures a local information need, matching it alone may overlook constraints in the original question $Q$. We therefore score documents against the composed query $\tilde{q}_t = Q \oplus q_t$, accounting for both local relevance and global constraints. To accommodate the System One model's bounded input and enable deeper reranking, we represent each document $d$ using a query-relevant passage of $m$ tokens. Rather than selecting the leading tokens, which may miss evidence deeper in the document, we select the span $w$ of $m$ tokens starting at the earliest occurrence of a non-stopword term from the agent-generated sub-query $q_t$. We use $q_t$ alone for passage selection to focus on the agent's immediate information need, while retaining $Q$ during reranking to account for relevance to the global information need. We then prepend the document title, yielding $\hat{d} = \mathrm{title}(d) \oplus w$. This compact representation prioritizes relevant evidence while allowing more candidates to fit within the model's input budget.

\begin{table}[t]
\centering
\small
\setlength{\tabcolsep}{2pt}
\caption{Effect of query and document representations on macro-averaged reranker recall and latency (ms).}
\vspace{-0.5em}
\label{tab:reranker_representation}
\scalebox{0.79}{
\begin{tabular}{l cccccccc @{\hskip 6pt} r}
\toprule
 & \multicolumn{4}{c}{\textbf{Gold}} & \multicolumn{4}{c}{\textbf{Evidence}} & \\
\cmidrule(lr){2-5}\cmidrule(lr){6-9}
 & \multicolumn{2}{c}{R@5} & \multicolumn{2}{c}{R@10} & \multicolumn{2}{c}{R@5} & \multicolumn{2}{c}{R@10} & Latency \\
\cmidrule(lr){2-3}\cmidrule(lr){4-5}\cmidrule(lr){6-7}\cmidrule(lr){8-9}
\textbf{Reranker} & Lead & Overlap & Lead & Overlap & Lead & Overlap & Lead & Overlap & (ms)$\downarrow$ \\
\midrule
BM25 & \multicolumn{2}{c}{26.3} & \multicolumn{2}{c}{34.4} & \multicolumn{2}{c}{27.6} & \multicolumn{2}{c}{33.8} & -- \\
\midrule
\multicolumn{10}{l}{\textit{Query: sub-query only}} \\
GTE-ModernBERT    & 29.4 & 31.2 & 36.5 & 39.0 & 27.1 & 27.6 & 32.8 & 34.6 & 416 \\
monoT5        & 28.6 & 28.8 & 34.8 & 36.7 & 25.6 & 25.6 & 30.9 & 32.7 & \underline{397} \\
Qwen3-Reranker-4B & 28.7 & 30.5 & 33.9 & 36.4 & 26.0 & 26.9 & 31.9 & 33.9 & 4,833 \\
gpt-oss-120b       & 35.6 & \underline{36.9} & \underline{40.2} & \textbf{42.8} & \underline{32.0} & \underline{32.9} & \textbf{38.1} & \textbf{39.5} & 52,188 \\
RankGPT \cite{rankgpt}        & \textbf{37.8} & \textbf{37.1} & \textbf{41.5} & \underline{42.3} & \textbf{33.4} & \textbf{33.3} & \underline{38.0} & \underline{39.1} & 8,167 \\
\textbf{SOARS (Ours)}             & \underline{35.8} & 36.6 & 40.1 & 41.2 & 30.8 & 31.7 & 37.6 & 38.2 & \textbf{241} \\
\midrule
\multicolumn{10}{l}{\textit{Query: original question + sub-query}} \\
GTE-ModernBERT    & 32.5 & 33.4 & 38.0 & 38.7 & 27.9 & 28.4 & 33.5 & 34.3 & 712 \\
monoT5                  & 27.6 & 27.8 & 33.6 & 33.5 & 23.1 & 23.1 & 29.8 & 30.1 & \underline{695} \\
Qwen3-Reranker-4B & 36.3 & 36.2 & 41.2 & 41.3 & 33.2 & 33.3 & 39.4 & 39.6 & 8,236 \\
gpt-oss-120b      & 42.6 & 43.3 & \underline{45.9} & 46.0 & 37.7 & 37.8 & 42.7 & 42.5 & 51,427 \\
RankGPT         & \textbf{43.6} & \underline{43.6} & \textbf{46.5} & \underline{46.6} & \textbf{39.2} & \underline{39.2} & \textbf{43.8} & \underline{43.6} & 8,706 \\
\textbf{SOARS (Ours)}             & \underline{42.9} & \textbf{44.0} & \underline{45.9} & \textbf{47.2} & \underline{38.9} & \textbf{39.4} & \underline{43.6} & \textbf{44.4} & \textbf{254} \\
\midrule
\textit{Oracle} & \multicolumn{2}{c}{46.3} & \multicolumn{2}{c}{48.0} & \multicolumn{2}{c}{41.6} & \multicolumn{2}{c}{45.5} & -- \\
\bottomrule
\end{tabular}%
}
\vspace{-1em}

\end{table}

\begin{table*}[t]
\centering
\small
\renewcommand{\arraystretch}{0.9}
\setlength{\tabcolsep}{3pt}
\setlength{\aboverulesep}{1pt}
\setlength{\belowrulesep}{3pt}
\newcolumntype{Q}{>{\centering\arraybackslash}p{3.4em}}
\newcolumntype{R}{>{\centering\arraybackslash}p{2.7em}}
\newcolumntype{W}{>{\centering\arraybackslash}p{4.6em}}
\newcolumntype{D}{>{\centering\arraybackslash}p{4.9em}}
\newcommand{\sig}{\rlap{$^{\dagger}$}}
\caption{Deep search agent performance across LLMs on BrowseComp-Plus and WebShaper. SOARS significantly improves accuracy, previewed recall, and behavior recall, exposing agents to more relevant evidence. Docs Read denotes the average number of documents read in full per agent-generated sub-query after previewing, with higher values indicating more evidence selected for deeper inspection. $^{\dagger}$ denotes significant improvement over the baseline under a paired two-tailed $t$-test ($p < 0.05$).}
\vspace{-1em}
\label{tab:main_results}
\scalebox{0.86}{
\begin{tabular}{l QQ RRRRRR D @{\hskip 1em} QQ WWW D}
\toprule
\multirow{3}{*}{\textbf{LLM}} & \multicolumn{9}{c}{\textbf{BrowseComp-Plus \cite{browscompplus}}} & \multicolumn{6}{c}{\textbf{WebShaper \cite{webshaper}}} \\
\cmidrule(lr){2-10}\cmidrule(lr){11-16}
  & \multicolumn{2}{c}{Answer Quality} & \multicolumn{2}{c}{Surfaced} & \multicolumn{2}{c}{Previewed} & \multicolumn{2}{c}{Behavior} & \multirow{2}{*}{Docs Read$\uparrow$} & \multicolumn{2}{c}{Answer Quality} & \multicolumn{3}{c}{Recall} & \multirow{2}{*}{Docs Read$\uparrow$} \\
\cmidrule(lr){2-3}\cmidrule(lr){4-5}\cmidrule(lr){6-7}\cmidrule(lr){8-9}\cmidrule(lr){11-12}\cmidrule(lr){13-15}
 & Acc. & Calib.$\downarrow$ & Gold & Evi. & Gold & Evi. & Gold & Evi. & & Acc. & Calib.$\downarrow$ & Surfaced & Previewed & Behavior & \\
\midrule
GPT-5.5          & 82.7 & 15.4 & 94.4 & 94.7 & 72.9 & 73.6 & 56.1 & 58.9 & 5.0 & 69.5 & 17.3 & 88.4 & 38.2 & 26.0 & 7.0 \\
\quad + \textbf{SOARS}     & \textbf{91.1}\sig & \textbf{4.3}\sig & \textbf{97.8} & \textbf{95.3} & \textbf{88.7}\sig & \textbf{83.3}\sig & \textbf{79.0}\sig & \textbf{68.7}\sig & \textbf{7.1}\sig & \textbf{77.0}\sig & \textbf{10.2}\sig & 86.9 & \textbf{47.9}\sig & \textbf{34.2}\sig & \textbf{10.5}\sig \\
\midrule
GPT-5.4 Mini     & 73.9 & 11.1 & 94.9 & 92.2 & 69.9 & 63.7 & 55.6 & 45.7 & 4.9 & 57.5 & 17.7 & 84.7 & 30.6 & 19.3 & 6.1 \\
\quad + \textbf{SOARS}     & \textbf{84.7}\sig & \textbf{8.7}\sig & \textbf{95.5} & 92.1 & \textbf{81.7}\sig & \textbf{74.8}\sig & \textbf{67.8}\sig & \textbf{54.9}\sig & \textbf{7.0}\sig & \textbf{64.5}\sig & \textbf{16.4} & 83.7 & \textbf{38.7}\sig & \textbf{24.6}\sig & \textbf{8.8}\sig \\
\midrule
Opus 4.7  & 70.4 & 10.3 & 86.8 & 81.2 & 52.5 & 43.3 & 43.0 & 30.4 & 2.1 & 67.5 & 4.4 & 68.8 & 21.4 & 14.4 & 3.2 \\
\quad + \textbf{SOARS}     & \textbf{82.7}\sig & \textbf{9.7} & 85.5 & 78.5 & \textbf{65.0}\sig & \textbf{53.9}\sig & \textbf{50.0}\sig & \textbf{35.3}\sig & \textbf{2.4}\sig & \textbf{68.0} & 5.9 & 64.3 & \textbf{28.9}\sig & \textbf{20.6}\sig & \textbf{4.3}\sig \\
\midrule
Haiku 4.5 & 54.9 & 17.1 & 94.1 & 60.1 & 58.2 & 46.3 & 40.9 & 39.4 & 8.8 & 48.5 & 16.2 & 87.4 & 35.5 & 23.2 & 8.8 \\
\quad + \textbf{SOARS}     & \textbf{73.9}\sig & \textbf{10.6}\sig & \textbf{96.1} & \textbf{94.7} & \textbf{79.3}\sig & \textbf{75.7}\sig & \textbf{62.3}\sig & \textbf{54.7}\sig & \textbf{10.2}\sig & \textbf{57.5}\sig & 16.3 & 87.1 & \textbf{46.2}\sig & \textbf{31.2}\sig & \textbf{11.6}\sig \\
\midrule
Gemini 2.5 Pro   & 30.1 & 36.5 & 73.1 & 71.5 & 26.4 & 25.5 & 16.8 & 14.3 & 1.2 & 38.0 & 30.7 & 71.9 & 16.7 & 6.8 & 1.1 \\
\quad + \textbf{SOARS}     & \textbf{49.3}\sig & \textbf{30.1}\sig & 72.6 & 68.4 & \textbf{45.4}\sig & \textbf{41.4}\sig & \textbf{28.6}\sig & \textbf{22.0}\sig & \textbf{1.7}\sig & \textbf{44.0} & 34.3 & 71.1 & \textbf{25.2}\sig & \textbf{10.7}\sig & \textbf{1.4}\sig \\
\midrule
Gemini 2.5 Flash & 31.9 & 20.1 & 80.3 & 79.1 & 34.8 & 34.6 & 21.9 & 19.9 & 2.4 & 27.5 & 30.1 & 77.5 & 19.5 & 8.1 & 1.7 \\
\quad + \textbf{SOARS}     & \textbf{47.5}\sig & \textbf{17.6} & \textbf{81.4} & 78.4 & \textbf{55.7}\sig & \textbf{51.5}\sig & \textbf{33.5}\sig & \textbf{28.4}\sig & \textbf{3.3}\sig & \textbf{35.5}\sig & \textbf{29.0} & 75.1 & \textbf{28.3}\sig & \textbf{12.7}\sig & \textbf{2.3}\sig \\
\bottomrule
\end{tabular}%
}
\vspace{-1em}
\end{table*}

Let $\mathcal{J}$ denote the System One reranker. We provide the top $k$ documents from $L_{q_t}$ in a single inference call to $\mathcal{J}$, which scores each query-document pair pointwise. It grades each candidate on an ordinal relevance scale of four levels, $\mathit{rel} \in \{0, 1, 2, 3\}$, corresponding to \emph{off-topic}, \emph{on a related topic}, \emph{partly supplies}, and \emph{fully supplies} the information needed to answer the query. Rather than committing to a single level, $\mathcal{J}$ returns a calibrated distribution $\mathbf{p}^{d}$ over the four levels for each document $d$, conditioned on the composed query $\tilde{q}_t$ and its passage $\hat{d}$. The probability assigned to each relevance level is defined as:
\begin{equation}
    p^{d}(\mathit{rel}) = \Pr\nolimits_{\mathcal{J}}\!\left(\mathit{rel} \mid \tilde{q}_t, \hat{d}\right),
\end{equation}
where $\Pr\nolimits_{\mathcal{J}}$ denotes $\mathcal{J}$'s calibrated probability and $\sum_{\mathit{rel}=0}^{3} p^{d}(\mathit{rel}) = 1$. We score each document by the expected relevance level under this distribution:

\begin{equation}
    s(d) = \mathbb{E}_{\mathit{rel} \sim \mathbf{p}^{d}}\!\left[\mathit{rel}\right] = \sum_{\mathit{rel}=0}^{3} \mathit{rel} \cdot p^{d}(\mathit{rel}).
\end{equation}

We then reorder the top $k$ documents in descending order of $s(d)$. Using the full distribution rather than the most likely level provides a graded ranking signal, while single-call scoring keeps the per-search overhead low.

\section{Experiments}
\label{sec:experiments}

\subsection{Setup}

\subsubsection{Datasets}
We evaluate SOARS on two benchmarks designed to assess deep search agents on complex, reasoning-intensive questions. \textbf{BrowseComp-Plus}~\cite{browscompplus} contains 830 complex questions and approximately 100K documents, with human-verified judgments distinguishing evidence documents needed to answer a question from gold documents that also contain the answer. \textbf{WebShaper}~\cite{webshaper} provides only gold documents. We randomly sample 200 questions and search the approximately 1M-document corpus released by \cite{agentir}.

\subsubsection{Reranker Baselines}
\label{sec:baselines}
We compare SOARS against monoT5~\cite{monobert} and gpt-oss-120b with medium reasoning~\cite{openai2025gptoss}, following their use in prior deep search agent work~\cite{meng2026ranking,rerankbeforeyoureason}. We also include RankGPT~\cite{rankgpt} with GPT-4.1 as its backbone, Qwen3-Reranker-4B~\cite{qwen3embedding}, and GTE-ModernBERT~\cite{gte}. We run the latter two rerankers, along with monoT5, on a single NVIDIA A10 GPU.

\subsubsection{Evaluation Metrics}
We organize our metrics around the three research questions. For \textbf{retrieval effectiveness} (RQ1), we measure the recall of the reranked list that each search exposes to the agent. For \textbf{agent effectiveness} (RQ2), following prior work~\cite{browscompplus,agentir,piserini,rerankbeforeyoureason}, we measure end-to-end Accuracy and Calibration Error with GPT-4.1 as the judge, and we trace gold/evidence through the trajectory with three recalls~\cite{piserini}. Specifically, \textit{surfaced recall} is computed over the documents returned by the agent's searches, \textit{previewed recall} over those whose snippets the agent inspects, and \textit{behavior recall} over those it opens or cites in its answer. For \textbf{agent efficiency} (RQ3), we report average search calls, per-query latency, and total cost. We calculate costs using submission-time API rates per million input, output, and cache-read tokens.

\subsubsection{Implementation Details}
We integrate SOARS inside the Pi-Serini agent framework~\cite{piserini} and study six LLM backbones of different families and sizes, namely GPT-5.5, GPT-5.4 Mini, Claude Opus 4.7, Claude Haiku 4.5, Gemini 2.5 Pro, and Gemini 2.5 Flash. Each backbone runs at medium reasoning effort under a 300-second per-query budget. Following~\cite{piserini}, we use BM25 ($k_1 = 25$, $b = 1$) as the first-stage retriever. For System One reranking, we employ Jev~\cite{jev}, reranking the top $k=100$ BM25 candidates at each search using document titles and query-relevant passages of $m=128$ tokens. Jev costs \$0.042 per million input tokens, with free output~\cite{jev}. For the vanilla baseline without SOARS, we reuse Pi-Serini's released trajectories for GPT-5.5, Opus 4.7, and Haiku 4.5 to evaluate answer accuracy using GPT-4.1 as the judge, while running the remaining backbones under identical settings.

\definecolor{citeblue}{HTML}{1A4FB0}

\begin{figure*}[t]
  \centering

    \includegraphics[width=0.97\textwidth,
            trim=0cm 0cm 0cm 0cm,
        clip
        ]{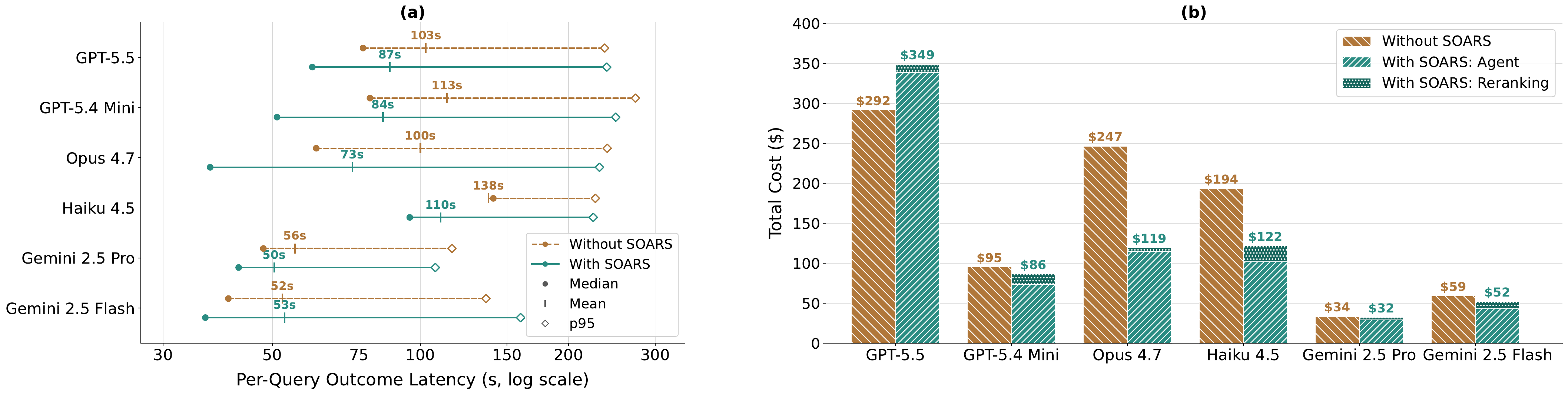}
\vspace{-0.5em}
\caption{Effect of SOARS on per-query latency (a) and total cost (b) of deep search agents across six different LLM backbones.}
\vspace{-1em}
\label{fig:efficiency}
\end{figure*}

\subsection{Results and Findings}

\subsubsection{RQ1: Retrieval Effectiveness}

We investigate how System One models can be adapted for in-loop reranking and compare their effectiveness against existing rerankers. We randomly select 50 BrowseComp-Plus questions comprising 1,044 sub-queries generated by GPT-5.4 Mini and their BM25 candidates. For each sub-query, SOARS and all baselines rerank the top 100 BM25-retrieved documents. We examine two query representations to assess global and local information needs in agentic search. The first uses the agent-generated sub-query to capture the immediate retrieval objective, while the second combines it with the original question to preserve broader context and constraints. We also compare two document representations, using either the leading tokens (\emph{Lead}) or a passage starting at the earliest occurrence of a non-stopword sub-query term (\emph{Overlap}), capturing evidence deeper in documents. These comparisons assess how query context and passage selection affect reranking effectiveness. Table~\ref{tab:reranker_representation} reports macro-averaged Recall@5, Recall@10, and reranking latency at the question level. The oracle prioritizes ground-truth gold or evidence documents within each BM25 candidate pool, establishing an upper bound on recall.

Both query and document representations influence retrieval effectiveness. Combining the original question with agent-generated sub-query improves recall for most rerankers, while \emph{Overlap} generally outperforms \emph{Lead}. Under the combined setting, SOARS achieves the highest recall across all four metrics, reaching 47.2\% gold and 44.4\% evidence Recall@10. These results approach the oracle upper bound, demonstrating that System One reranking can be highly effective when jointly considering global and local information needs through focused document passages.

Beyond retrieval effectiveness, SOARS offers a substantial latency advantage. Although GTE-ModernBERT and monoT5 offer low per-query latency, SOARS achieves substantially higher recall with even lower latency. More notably, using overlap-selected passages and the original question combined with the sub-query, SOARS consistently outperforms RankGPT, the strongest LLM-based reranker, while requiring only 254 ms per query compared to 8.7 seconds. This combination of strong retrieval effectiveness and low latency makes System One reranking particularly suitable for repeated invocation throughout the agent's search trajectory.

\subsubsection{RQ2: Agent Effectiveness}

We assess how integrating System One reranking into deep search agents affects end-to-end answer accuracy and evidence recall across two benchmarks.

\textbf{BrowseComp-Plus.} SOARS significantly improves answer accuracy across all LLMs, with relative gains ranging from 10.2\% for GPT-5.5 to 63.8\% for Gemini 2.5 Pro. These improvements are particularly pronounced for weaker agents, narrowing the performance gap with stronger backbones. Calibration error also consistently decreases, indicating more reliable confidence estimates. More importantly, the evidence recall analysis reveals where these gains originate. Despite using the same first-stage retriever and achieving similar surfaced gold recall, SOARS significantly improves previewed and behavior recall across all LLMs. For GPT-5.5, previewed gold recall increases from 72.9 to 88.7, while behavior gold recall rises from 56.1 to 79.0.

These findings suggest that SOARS improves answer accuracy by promoting already-retrieved evidence into the documents agents preview, read, and cite. Furthermore, the average number of documents read increases significantly across all LLMs. As agents preview short document excerpts before selectively opening promising ones for full-text reading, this suggests that SOARS helps identify more useful evidence for deeper inspection. Notably, these gains are accompanied by fewer search calls (Figure~\ref{fig:motivation_figure}), suggesting that better evidence prioritization reduces the need for additional retrieval.

\textbf{WebShaper.} The same pattern generalizes to WebShaper, where previewed and behavior recall increase significantly for every backbone, with significant accuracy improvements for most. The number of documents read also increases significantly across all backbones, suggesting more promising evidence is selected for deeper inspection. These findings confirm that SOARS consistently improves evidence utilization throughout the search trajectory, contributing to higher answer accuracy.

\subsubsection{RQ3: Agent Efficiency}

We evaluate the efficiency of per-step System One reranking across three dimensions, namely the number of search calls, per-query latency, and cost. We report results on BrowseComp-Plus, with broadly similar trends observed on WebShaper.

\textbf{Search calls.} SOARS reduces search calls across all LLMs (Figure~\ref{fig:motivation_figure}), especially for search-intensive agents, from 23 to 14 for GPT-5.4 Mini and 32 to 21 for Haiku 4.5. By promoting relevant documents to the top of the ranked list, SOARS helps agents identify and utilize useful evidence earlier in the search trajectory, reducing the need for subsequent searches.

\textbf{Latency.} Despite adding a reranking call per search, SOARS reduces per-query latency for nearly all backbones (Figure~\ref{fig:efficiency}a), from 113 to 84 seconds for GPT-5.4 Mini and 100 to 73 seconds for Opus 4.7. Fewer search calls offset the reranker's modest overhead, yielding a net latency reduction. These findings demonstrate that lightweight reranking can accelerate the overall search process by reducing the time spent on repeated search interactions.

\textbf{Cost.} Total cost decreases for five of six backbones (Figure~\ref{fig:efficiency}b), particularly for expensive agents, from \$247 to \$119 for Opus 4.7 and \$194 to \$122 for Haiku 4.5. Reduced search effort offsets the minimal reranking cost, lowering overall agent compute. GPT-5.5 is the sole exception, with a modest cost increase potentially due to reading more evidence, contributing to higher accuracy (Table~\ref{tab:main_results}).

\section{Concluding Remarks}
\label{sec:conclusion}

We introduced SOARS, a training-free framework that adapts a System One model as a reranker inside the deep search agent's iterative loop. Across six LLM backbones on BrowseComp-Plus and WebShaper, SOARS consistently improves answer accuracy and evidence recall while reducing the agent's search calls, latency, and cost. These results show that fast, calibrated System One models make per-step reranking both effective and efficient, establishing them as a practical component for building deep search agents.

\newpage
\bibliographystyle{ACM-Reference-Format}
\balance
\bibliography{references}

\end{document}